\documentstyle[preprint,revtex]{aps}
\begin{document}
\newcommand{\Tr}{\mbox{Tr}}
\newcommand{\Jpsi}{J\!/\!\psi}
\newcommand{\slh}{\hspace{-.5em}/}
\newcommand{\mpsi}{m_\psi}
\newcommand{\oas}{\mbox{$\mbox{O}(\alpha_{s})$}}
\newcommand{\oasz}{\mbox{$\mbox{O}(\alpha_{s}^{2})$}}
\newcommand{\as}{\mbox{$\alpha_{s}$}}
\newcommand{\asz}{\mbox{$\alpha_{s}^{2}$}}
\newcommand{\lms}{\mbox{$\Lambda_{\overline{\mbox{\tiny MS}}}$}}
\def\Lms#1{\Lambda_{\overline{MS}}^{(#1)}}
\def\asp{{\alpha_s\over\pi}}
\def\ams{\alpha_{\overline {MS}} }
\font\fortssbx=cmssbx10 scaled \magstep2
\hbox to \hsize{
\setlength{\footskip}{1.5cm}
\frenchspacing
\everymath={\displaystyle}
  \def\thebibliography#1{{\bf{References}}\list
 {[\arabic{enumi}]}{\settowidth\labelwidth{[#1]}\leftmargin\labelwidth
   \advance\leftmargin\labelsep
   \usecounter{enumi}}
   \def\newblock{\hskip .11em plus .33em minus -.07em}
   \sloppy
   \sfcode`\.=1000\relax}
  \let\endthebibliography=\endlist
\hskip.5in \raise.1in\hbox{\fortssbx University of Wisconsin - Madison}
\hfill$\vcenter{\hbox{\bf MAD/PH/822}
            \hbox{\bf YUMS 94--11}
            \hbox{\bf SNUTP 94--34}
            \hbox{(July 1994)}}$ }

\vspace{.5in}
\begin{title}
Testing $\Jpsi$ Production and Decay Properties\\
in Hadronic Collisions
\end{title}
\author{C.S.Kim$^a$ and E. Mirkes$^{b}$}
\begin{instit}
$^a$ Department  of Physics, Yonsei University, Seoul 120--749, KOREA\\
$^b$Physics Department, University of Wisconsin, Madison WI 53706, USA\\
\end{instit}
\begin{abstract}
The polar and azimuthal angular distributions for the lepton pair
arising from the decay of a $\Jpsi$ meson produced at transverse momentum
$p_{_T}$  balanced by a photon [or gluon] in hadronic
collisions are calculated in the color singlet model (CSM).
It is shown that the general structure of the decay lepton distribution
is controlled by four invariant structure functions, which are functions
of the transverse momentum and the rapidity of the $\Jpsi$.
We found that two of these structure functions
[the longitudinal and transverse interference structure functions]
are identical in the CSM.
Analytical and numerical results are given  in the Collins-Soper
and in the Gottfried-Jackson frame.
We present a Monte Carlo study of the effect of acceptance cuts
applied to the leptons and the photon for $\Jpsi+\gamma$ production
at the Tevatron.
\end{abstract}

\thispagestyle{emtpy}
\newpage
\section{INTRODUCTION}
The study of the  properties of the bound states of heavy quarks
plays a central role in the understanding of
perturbative Quantum Chromodynamics. Besides of the intrinsic interest
in the bound state properties of a $\Jpsi$, the study of the
${}^3S_1- c\bar{c}$ bound state
is also of  particular importance in the measurement of $b$ quark
production.
So far there has been intensive experimental studies of the $\Jpsi$
production rate and transverse momentum distributions
in hadronic collisions both
at UA1 \cite{ua1} and CDF \cite{cdf}.
However, the observed $\Jpsi$ rate was found to be markedly higher
than the predicted one's in \cite{theo}, where
in addition to the direct charmonium production also contributions from
the $\chi$ and the production resulting from $B$ decays were
taken into account.
It has recently been pointed out in \cite{eric} that at large
transverse momentum of the $\Jpsi$  an additional production mechanism
comes from fragmentation contributions of a gluon or a charm quark
into charmonium states.

In this paper we propose to measure the  decay lepton distribution
of the $\Jpsi$ as a detailed test of the production mechanism of the
$c\bar{c}$ bound state.
We restrict ourself in the present paper to the study of $\Jpsi$ produced
in association with a photon. This has several advantages.
First of all the experimental signature of a $\Jpsi$ [decaying into
an $e^-e^+$ or $\mu^-\mu^+$ pair] and a $\gamma$ with balancing
transverse momentum is a very clean final state.
Second, there are no fragmentation  contributions or contributions
from radiative $\chi$ decays\footnote{
The radiative $\chi_J$ decays can produce
$J/\psi$ at both low and high $p_{_T}$, but the photon produced will be
soft [$E \sim {\cal O}(400\;{\rm MeV})$].}
to $\Jpsi + \gamma$ production and the only subprocess
contributing to a  $\Jpsi+\gamma$ final state is
the gluon gluon fusion process.

For $\Jpsi$'s produced with
transverse momentum $p_{_T}$ [balanced by the additional photon] one
can define an event plane spanned by the beam and the
$\Jpsi$ momentum direction which provides
a reference plane for a detailed study of angular correlations.
The decay lepton distribution in $\Jpsi\rightarrow l^- l^+$
in the $\Jpsi$ rest frame is determined by the polarization of the $\Jpsi$.
Therefore, the study of the angular distribution can be used as an analyzer
of the $\Jpsi$-polarization.
It is thus possible to test the underlying $\Jpsi$-production dynamics
[in our case the color singlet model (CSM) \cite{csm}]
in much more detail than is possible by rate measurements alone.
It is shown that the  angular distribution of the
leptons in the $\Jpsi$-rest frame
 has the general form
\begin{equation}
\sigma \sim (1+\cos\theta^2)
  + \frac{1}{2} \, A_{0} \, (1-3\cos^{2}\theta)
  +  A_{1}  \, \sin 2\theta \cos\phi
  + \frac{1}{2} \, A_{2}  \, \sin^{2}\theta\cos 2\phi\, ,
\label{a}
\end{equation}
where $\theta$ and $\phi$ denote the polar and azimuthal angle of the
decay leptons in the $\Jpsi$-rest frame.
The coefficients $A_i$ in Eq.~(\ref{a}) are functions of the
transverse momentum and rapidity of the $\Jpsi$.
We present analytical results for these coefficients
in the Collins Soper (CS) \cite{cs} and Gottfried Jackson (GJ) frame.
The functions $A_0$ and $A_2$
are found to be equal in the CSM. An experimental
analyses  of this equality would be a sensitive test of the CSM.
We show numerical results for the transverse momentum [$p_T(\Jpsi)$]
and rapidity [$y(\Jpsi)$] dependence of the coefficients $A_i$
and study the effect of the acceptance cuts on the
$\phi$ and $\cos\theta$ distributions in the CS and
GJ frame.

The coefficients $A_i$
 are ratios of helicity cross sections
[see Eq.~(\ref{aintr})]
and the  uncertainties from higher order QCD corrections
and the wave function of the bound state are expected to cancel
in this ratio.
This has been shown for the related process of high $p_T$
lepton pair production in the Drell-Yan process \cite{npb}, where large
QCD corrections to the individual helicity cross sections
almost cancel in the corresponding ratios $A_i^{DY}$.
A similar effect is expected in the case of $\Jpsi$ production and
our LO predictions for the $A_i$
and the [normalized] $\phi$ and $\cos\theta$ distributions
provide therefore a useful test of $\Jpsi$ production.

Our analytical results for the coefficients $A_i$
are also valid for $\Jpsi+g$ production
within the CSM.
As mentioned before, there are additional contributions to
this final state \cite{eric}. A study of the angular distribution
as proposed in this paper
may allow to disentangle the different contributions.

Recently, $\Jpsi$'s produced in association with a photon  has been
proposed as a clean probe to study the gluon distribution 
and the
polarized gluon distribution of proton 
as well as  to investigate
qualitatively
the production
mechanism of heavy quark bound-states\cite{kim1}.
An investigation of the decay lepton distribution and
structure functions of $\Jpsi$'s production and decay
in $e^-e^+$--experiments is presented in \cite{volker}.

The remainder of this paper is organized as follows:
In Sec.~II, analytical results for the helicity cross sections
and the coefficients $A_i$ are derived in the CSM
for the CS
and GJ frame and
the formalism
for describing the angular distributions is discussed.
Numerical results are presented in Sec.~III and  summary remarks are
given in Sec.~IV. Some technical details of the calculation are
relegated to the appendix.

\section{STRUCTURE FUNCTIONS AND ANGULAR DISTRIBUTIONS}
We consider the angular distribution of the leptons coming from the
leptonic decay of $\Jpsi$'s produced with non-zero transverse
momentum in association with a photon
in high energy proton-antiproton collisions. For definiteness
we take
\begin{equation}
p(P_{1}) + \bar{p}(P_{2}) \rightarrow \Jpsi(P) +\gamma(k) +X
\rightarrow l^-(l)+l^+(l') + \gamma(k)+ X \>,
\end{equation}
where the quantities in the parentheses denote the
four-momenta.
In  leading order of perturbative QCD [$O(\alpha_s^2)$],
 $\Jpsi+\gamma$  can only be produced in $gg$
fusion:
  \begin{equation}
g(p_{_1})+g(p_{_2})\rightarrow \Jpsi(P)+ \gamma(k) +X\>.
\label{partonic}
  \end{equation}

In the parton model the hadronic cross section is obtained by folding
the hard parton level cross section with the respective parton
densities:
\begin{equation}
\frac{d\sigma^{h_{1}h_{2}}}{ dp_{_T}^{2}\,dy\,d\Omega^{\ast} } =
\int\,\,dx_{1}dx_{2}
g^{h_{1}}(x_{1},\mu_F^{2})
g^{h_{2}}(x_{2},\mu_F^{2})
\,
\frac{{s}\,d{\hat{\sigma}}}{ d{t}\,d{u}\,d\Omega^{\ast}}
\left( x_{1}P_{1},x_{2}P_{2},\as(\mu_R^{2}) \right) \>,
\label{wqhad}
\end{equation}
where $g^h(x,\mu_F^2)$
is the probability density to find a gluon  with fraction $x$
in hadron $h$ if it is probed at a scale $\mu_F^2$.
The partonic cross section for the process in Eq. (\ref{partonic})
is denoted by $d{\hat{\sigma}}$.
Denoting hadron level and parton
level quantities by upper and lower case characters, respectively,
the hadron and parton level Mandelstam variables are  defined by
\begin{equation}
S = (P_1 + P_2)^2 \>, \qquad T = (P_1 - P)^2 \>, \qquad U = (P_2 - P)^2 \>,
\end{equation}
and
  \begin{equation}
  \begin{array}{lclcl}
\label{kleinmandeldef}
s&=&(p_{_1}+p_{_2})^{2}&=&x_{1}x_{2}S \>,\\[2mm]
t&=&(p_{_1}-P)^{2}    &=&x_{1}(T-P^2)+P^2\>,\\[2mm]
u&=&(p_{_2}-P)^{2}    &=&x_{2}(U-P^2)+P^2\>,
  \end{array}
\end{equation}
where $p_1 = x_1 P_1$ and $p_2 = x_1 P_2$. The rapidity $y$ of the
$\Jpsi$ in the laboratory frame can be written as
\begin{equation}
y=\frac{1}{2}\log\left(\frac{P^2-U}{P^2-T}\right)\>,
\end{equation}
and the transverse momentum of the $\Jpsi$ is related to the
Mandelstam variables via
\begin{equation}
p_{_T}^{2}=\frac{(P^2-U)(P^2-T)}{S}-P^2=\frac{ut}{s}\>.
\end{equation}.

The angles $\theta$ and
$\phi$ in $d\Omega^{\ast} = d\cos\theta\,d\phi$ in Eq. (\ref{wqhad})
are the polar and azimuthal
decay angles of the leptons  in the $\Jpsi$ rest frame with respect to
a coordinate system  described below.
The angular dependence in  (\ref{wqhad}) can be extracted by
introducing four helicity cross sections
corresponding to the four none zero combinations of
polarization density matrix elements [see Appendix A of this paper,
and Appendix C of \cite{npb} for details]:
\begin{eqnarray}
\frac{16\pi}{3}\frac{d{\sigma}}{dp_{_T}^2\,dy\,d\cos\theta\,d\phi}
&=& \frac{d{\sigma}^{U+L}}{dp_{_T}^2\,dy}\,(1+\cos^2\theta)
\,+\, \frac{d{\sigma}^{L}}{dp_{_T}^2\,dy}\,(1-3\cos^2\theta)
\label{hadronwinkel}
\\
&+& \frac{{d\sigma}^{T}}{dp_{_T}^2\,dy}\,2\sin^2\theta\cos2\phi
\,+\, \frac{d{\sigma}^{I}}{dp_{_T}^2\,dy}\,
2\sqrt{2}\sin2\theta\cos\phi \nonumber\>.
\end{eqnarray}
The hadronic  helicity cross sections
$\frac{d{\sigma}^{\alpha}}{dp_{_T}^2\,dy}$
are again obtained by convoluting the partonic helicity  cross sections
with the gluon densities. One has
\begin{equation}
\frac{d\sigma^{\alpha}}{ dp_{_T}^{2}\,dy } =
\int\,\,dx_{1}dx_{2}
g^{h_{1}}(x_{1},\mu_F^{2})
g^{h_{2}}(x_{2},\mu_F^{2})
\,
\frac{{s}\,d{\hat{\sigma}^\alpha}}{ d{t}\,d{u}}
\label{wqalphahad}\>.
\end{equation}
Each of the partonic  helicity cross sections is calculated
in the CSM.
The unpolarized differential production cross section is denoted
by $\hat{\sigma}^{U+L}$ whereas $\hat{\sigma}^{L,T,I}$ characterize the
polarization of the $\Jpsi$, i.e.
the  cross section for the longitudinal polarized $\Jpsi$'s is denoted
by $\hat{\sigma}^L$, the transverse-longitudinal
interference by $\hat{\sigma}^I$,
and the transverse interference
by $\hat{\sigma}^T$
(all with respect to the $z$-axis of the chosen $\Jpsi$ rest frame).
The results for the helicity cross sections $\hat{\sigma}^{\alpha}$
are dependent of the choice of the
$z$ axis in the rest frame of the $\Jpsi$. We will derive
explicit results for the
Collins-Soper (CS) and the Gottfried-Jackson (GJ) frame.
The Lorentz transformation matrix for the  boost
from the laboratory frame to the CS  and GJ frame is given in Appendix B.

In the CS frame \cite{cs} the $z$-axis
bisects the angle between ${\vec{P}_{1}}$ and ${-\vec{P}_{2}}$
\begin{equation}
    \begin{array}{ll}
CS: \,\,\,& {\vec{P}_{1}} =
\,\,\, E_{1}\,\, (\sin\gamma_{CS},0,\cos\gamma_{CS})\>,   \\[1mm]
          & {\vec{P}_{2}} =
\,\,\,     E_{2}\,\,(\sin\gamma_{CS},0,-\cos\gamma_{CS})\>,
          \end{array}
\label{csdef}
\end{equation}
with
\begin{equation}
\cos\gamma_{CS}=
\sqrt{\frac{\mpsi^2 S}{(T-\mpsi^2)(U-\mpsi^2)}}
=\sqrt{\frac{\mpsi^2}{\mpsi^2+p_{_T}^2}}\>,
\label{winkelcsdef}
\end{equation}
\begin{equation}
\sin\gamma_{CS}=-\sqrt{1-\cos^2\gamma_{CS}}\>.
\end{equation}

In the GJ frame
(also known as $t$-channel helicity frame)
the $z$-axis is chosen parallel to the beam axis
\begin{equation}
          \begin{array}{ll}
GJ: \,\,\,& {\vec{P}_{1}} = E_{1} \,\,\,
                            (0,0,1)\>,\\[1mm]
          & {\vec{P}_{2}} = E_{2}\,\,\,
                            (\sin\gamma_{GJ},0,\cos\gamma_{GJ})\>,
          \end{array}
\label{gjdef}
\end{equation}
with
\begin{equation}
\cos\gamma_{GJ}=
 {1-\frac{2\mpsi^2 S}{(T-\mpsi^2)(U-\mpsi^2)}}
=\frac{p_{_T}^2-\mpsi^2}{p_{_T}^2+\mpsi^2}\>,
\end{equation}
\begin{equation}
\sin\gamma_{GJ}=-\sqrt{1-\cos^2\gamma_{GJ}}\>.
\end{equation}
The beam and target energies in the rest frame are
$$E_{1}=(\mpsi^2-T)/(2\mpsi), \,\,\,\, E_{2}=(\mpsi^2-U)/(2\mpsi)~.$$

The partonic helicity cross sections in Eq. (\ref{wqalphahad})
are calculated applying the technique described in \cite{npb}
to the CSM [see Appendix A for details].
They are given by:
\begin{equation}
\frac{s\hat{\sigma}^{\alpha}_{CS, GJ}}{dt du}
=\frac{16\pi \alpha\alpha_s^2 \mpsi}{27 s }
|R(0)|^2\,\,H^{\alpha}_{CS, GJ}\,
\delta(s+t+u-\mpsi^2)\>,
\label{partondef}
\end{equation}
with
\begin{eqnarray}
H_{CS}^{U+L}&=&
       \frac{s^2}{(t-\mpsi^2)^2(u-\mpsi^2)^2}
     + \frac{t^2}{(u-\mpsi^2)^2(s-\mpsi^2)^2}
     + \frac{u^2}{(s-\mpsi^2)^2(t-\mpsi^2)^2}\>,\hspace{1cm}\label{firstcs}
\\[2mm]
H_{CS}^{L}&=&
             \frac{ ut}{2(t-\mpsi^2)(u-\mpsi^2)}\left(
         \frac{4\mpsi^2s^3}{(s-\mpsi^2)^2(t-\mpsi^2)^2(u-\mpsi^2)^2}
          + H^{U+L}_{CS}\right)\>,\\[2mm]
H_{CS}^{T}&=&\frac{1}{2} H_{cs}^{L}\>,\\[2mm]
H_{CS}^{I}&=&
\frac{ \mpsi\sqrt{sut}\,\,\,s(s^2-ut)(t-u)}{
        \sqrt{2}(s-\mpsi^2)^2(t-\mpsi^2)^3(u-\mpsi^2)^3}\>,
\label{lastcs}
\end{eqnarray}
and for the GJ frame
\begin{eqnarray}
H_{GJ}^{U+L}&=& H_{CS}^{U+L}\>,\hspace{12cm}\label{firstgj}\\[2mm]
H_{GJ}^{L}&=&
 \frac{2\mpsi^2 s t u
(s^2+u^2)}{(s-\mpsi^2)^2(t-\mpsi^2)^4(u-\mpsi^2)^2}\>,\\[2mm]
H_{GJ}^{T}&=&\frac{1}{2} H_{GJ}^{L}\>,\\[2mm]
H_{GJ}^{I}&=&
-\frac{\mpsi\sqrt{ s t u }\,(s-u)[s^2(u-t)+u^2(s-t)]}{
        \sqrt{2}(s-\mpsi^2)^2(t-\mpsi^2)^4(u-\mpsi^2)^2}\>.
 \label{lastgj}
\end{eqnarray}
As mentioned before, $H^{U+L}$ denotes the matrix element contribution
for the production rate and was first calculated in
\cite{berger}. All other matrix elements correspond to the production of
{\it polarized} $\Jpsi$'s and are given here for the first time.
Replacing  $16\alpha$  by $15\alpha_s$ in Eq.~(\ref{partondef}),
the results in
Eqs.~(\ref{partondef}-\ref{lastgj}) are also valid for $\Jpsi+g$
production within the CSM.

Note, that the matrix elements can be expressed in terms of $p_{_T}^{2},y$
by noting that
\begin{eqnarray}
t&=&\mpsi^{2}-\sqrt{(\mpsi^{2}+p_{_T}^{2})S}\,\,x_{1}e^{-y}\nonumber \>,\\
u&=&\mpsi^{2}-\sqrt{(\mpsi^{2}+p_{_T}^{2})S}\,\,{x}_{2}e^{ y}\>,\label{tus}
\end{eqnarray}
so that e.g.
\begin{equation}
\frac{(\mpsi^2-t)(\mpsi^2-u)}{s}=p_{_T}^2+\mpsi^2 \>,\hspace{5mm}
\mbox{and also} \hspace{5mm} \frac{ut}{s}=p_{_T}^2\>.
\end{equation}
{}From the explicit expressions in Eqs.~(\ref{firstcs}-\ref{lastgj})
one observes, that the longitudinal and the transverse interference
cross sections $\sigma^{L}$ and $\sigma^{T}$  are related by
$\sigma_L=2\sigma_T$ and that they
vanish for $p_{_T}\rightarrow 0$ with $p_{_T}^2$.
The longitudinal-transverse interference cross section $\sigma^{I}$ vanishes
with $p_{_T}$, whereas the production cross section $\sigma^{U+L}$ for
unpolarized $\Jpsi$'s is finite in  the $p_{_T}\rightarrow 0$ limit.

Introducing standard angular coefficients \cite{cs}
\begin{equation}
A_{0}=\frac{2\,\, d\sigma^{L}}{d\sigma^{U+L}}\>,\hspace{1cm}
A_{1}=\frac{2\sqrt{2}\,\, d\sigma^{I}}{d\sigma^{U+L}}\>,\hspace{1cm}
A_{2}=\frac{4 \,\,d\sigma^{T}}{d\sigma^{U+L}}\>,\hspace{1cm}
\label{aintr}
\end{equation}
the angular distribution of Eq. (\ref{hadronwinkel}) is conveniently written
as:
\begin{eqnarray}
\frac {d\sigma}{d p_{_T}^{2}\,dy\, d\cos\theta \,d\phi}
&=&  \frac{3}{16\pi}\,
\frac{d\sigma^{U+L}}{ d p_{_T}^{2}\,dy}\,\,
           \,\left[  (1+\cos^{2}\theta) \nonumber
               +\,\, \frac{1}{2}A_{0} \,\,\, (1-3\cos^{2}\theta)
 \right.\\[2mm]
&&   \left.
\hspace{1.5cm} +  \,\,   A_{1}  \,\,\,\sin 2\theta \cos\phi \,\,
\,\, + \,\,   \frac{1}{2}A_{2}
\,\,\,\sin^{2}\theta\cos 2\phi    \right]\>.
  \label{ang}
\end{eqnarray}
Integrating the angular distribution in Eq. (\ref{ang}) over the azimuthal
angle
$\phi$ yields:
\begin{equation}
\frac{d\sigma}{dp_{_T}^{2}\,dy\,d\cos\theta}=C\,
                  (1+
                \alpha
\cos^{2}\theta)\>,
\label{alphadef}
\end{equation}
where
\begin{equation}
C= \,\frac{3}{8}\,
\frac{d\sigma^{U+L}}{dp_{_T}^{2}\,dy}
\left[1+\frac{A_{0}}{2}\right]\hspace{1cm}
\alpha=\frac{2-3A_{0}}{2+A_{0}}\>.
\label{alphaidef}
  \end{equation}
And integration over $\theta$ yields
\begin{equation}
\frac{d\sigma}{dp_{_T}^{2}\,dy\,d\phi}=
\,\frac{1}{2\pi}\,
\frac{d\sigma^{U+L}}{dp_{_T}^{2}\,dy}
         (1+
\beta\cos 2\phi )\>,
\label{betadef}
\end{equation}
where
\begin{equation}
\beta=\frac{A_{2}}{4}\>.\hspace{1cm}
\label{betaidef}
\end{equation}
Before discussing numerical results, let us briefly discuss  a possibility
strategy for extracting
angular coefficients.
By taking  moments with respect to an appropriate product of trigonometric
functions it is possible to disentangle the coefficients $A_i$.
A convenient definition of the moments through
\begin{equation}
\langle m \rangle =
\frac{\int d\sigma(p_T,y,\theta,\phi)\,\, m \,\,{d\cos\theta}\,
{d\phi}}{
\int d\sigma(p_T,y,\theta,\phi)\,\,{d\cos\theta}\,{d\phi}}\>,
\label{momentdef}
\end{equation}
which leads to the following results
\begin{eqnarray}
\langle 1 \rangle &=& 1 \>,\\[2mm]
\langle \frac{1}{2}(1-3\cos^2\theta ) \rangle &=&
\frac{3}{20}\,\,\left( A_0-\frac{2}{3}\right)\>, \\[2mm]
\langle \sin 2\theta\cos\phi  \rangle &=& \frac{1}{5} \,\,A_1 \>,\\[2mm]
\langle \sin^2\theta\cos2\phi  \rangle &=&\frac{1}{10}\,\, A_2\>.
\end{eqnarray}

\section{NUMERICAL RESULTS}
We will now present numerical results for $\Jpsi+\gamma$ production  at
the Tevatron collider center of mass energy [$\sqrt{S}=1.8$ TeV] including
the decay  $\Jpsi\rightarrow 
\mu^-\mu^+$.
All results are obtained using the  gluon density
parametrization from  GRV \cite{grv}
with $\Lms{4}=200$ MeV
and the one-loop formula for $\alpha_s$ with 4 active flavours.
If not stated otherwise,
we identify the renormalization scale $\mu_R^2$
and the factorization scale $\mu_F^2$ in Eq. (\ref{wqhad}) and set
them equal to
$\mu^2=\mu_F^2=\mu_R^2=(\mpsi^2+p_{_T}^2(\Jpsi))$.
The value for the bound state wave function at the origin $|R(0)|^2$
is determined from the leptonic decay width of $\Jpsi$:
$\Gamma (\Jpsi \rightarrow e^- e^+) = 4.72 $ keV, therefore
$|R(0)|^2 = 0.48$  GeV$^3$.

Fig.~1 shows the transverse momentum distribution of the
the $\Jpsi$ where we have applied the  cuts  $|y(\gamma,l^-,l^+)|<2.5$
and $p_T(l^-,l^+)>1.8$ GeV on the final state particles.
The $p_{_T}$ distributions  are leading order predictions
and are therefore strongly dependent on the choice of the scales $\mu^2$
and the radial wave function of the bound state.
In fact, the NMC collaboration found \cite{nmc} that the LO
predictions describes the shape of all kinematical variables quite well,
while the predictions for the normalization of the signal was too small
by a factor 2-3. This large ``K-factor'' is probably  due
to the nonrelativistic treatment of the $\Jpsi$
and due to higher order QCD corrections in the CSM.
One can  expect a
similar K-factor for our reaction, so that our result of Fig.~1
should be considered as conservative estimates.
To give a feeling for the scale dependence we  show results for
$\mu^2=\mu_F^2=\mu_R^2=(\mpsi^2+p_{_T}^2(\Jpsi))$ [solid line],
$\mu^2=(\mpsi^2+p_{_T}^2(\Jpsi))/4$ [dashed line], and
$\mu^2=4(\mpsi^2+p_{_T}^2(\Jpsi))$ [dotted line].

Fig.~2 shows the rapidity distribution of the $\Jpsi$
normalized to the total production cross section.
The following cuts are applied: 
$p_T(\gamma,l^-,l^+)>3$ GeV.
The shape of  this distribution is quite sensitive
to the  gluon density function, and therefore
useful to extract the gluon density  at the small $x$ \cite{kim1}.

In Fig.~3, we show the transverse momentum  distributions
of the  decay leptons from the $\Jpsi$ after the  cuts:
$|y(\gamma,l^-,l^+)|<2.5, \,\,p_T(l^-,l^+)>1.8$ GeV
and $p_T(\gamma)> 2$ GeV. We also plot the
spectrum of the harder (denoted by 'b')
and softer (denoted by 's') lepton separately, where we did not
distinguish between the charge of the leptons.
Note that together with the hard
photon, this harder lepton (denoted by 'b') can be used to construct
a trigger for our reaction.
Like in Fig.~1, the $p_T$ distributions of the leptons are
strongly dependent on the choice of the scales $\mu^2$
and the radial wave function of the bound state.

In Figs.~4 and 5 we show
numerical results for the coefficients $A_i$ in Eq.~(\ref{ang})
as a function of $p_T(\Jpsi)$ and $y(\Jpsi)$
 in the CS [Figs.~4,5(a)] and GJ [Figs.~4,5(b)] frame.
These coefficients
have been extracted from the Monte Carlo program  by using the moments
defined in Eq.~(\ref{momentdef}).
No acceptance cuts  have been applied to the leptons or the photon.
One observes that the coefficients are  strongly dependent
on $p_T(\Jpsi)$ both in the CS and GJ frame.
As  mentioned before, the coefficients $A_0$ and $A_2$ are exactly equal
in lowest order in both lepton pair rest frames.
The angular coefficient $A_1$ is zero in the CS frame
for all values of $p_{_T}(\Jpsi)$.
The reason is that the matrix element
for $A_1$ is antisymmetric in $u$ and $t$ [see Eqs. (\ref{lastcs},\ref{aintr})]
and therefore in $x_1$ and $x_2$, whereas
the product of the gluon distributions is symmetric under the interchange
of $x_1$ and $x_2$.
As a consequence, the rapidity distribution for $A_1$ in the
CS frame [Fig.~5(a)] is also antisymmetric around $y=0$.
However, this is  different   for the GJ frame [see Eq. (\ref{lastgj})].
All  coefficients $A_i$ vanish in the limit
$p_{_T}(\Jpsi)\rightarrow 0$, which can be
directly seen from our analytical expressions in Eqs.
(\ref{firstcs}-\ref{lastgj},\ref{aintr}).

A similar relation $A_0^{DY}=A_2^{DY}$
was found in LO [$O(\alpha_s)$] in the Drell-Yan process \cite{tung}
$p+\bar{p}\rightarrow V + X \rightarrow l^+l'+X$, where $V$ denotes a
gauge boson produced at high $p_T$.
In \cite{npb}, the complete NLO corrections to the coefficients
$A^{DY}_i$ are calculated and  the corrections
are found to be fairly small for the ratio's $A_i$ of the corresponding
helicity cross section.
We expect also here, that the  LO results
for the ratio's $A_i$ in $\Jpsi$ production
are almost not affected by higher order
QCD corrections. Note also, that the coefficients $A_i$
are not dependent on the bound state wave function.
The measurement of these coefficients
would be a sensitive  test of the production mechanism of $\Jpsi$'s.
However, as we will see later,
experimental cuts introduce additional complicated angular effects
and the resulting data sample can no longer be described by the simple
angular distribution in Eq.~(\ref{ang}).

Figures 6-11 show the normalized $\phi$ and $\cos\theta$ distributions
for the leptons in the CS and GJ frame
for three different bins in  the transverse momentum
of the $\Jpsi$.
To demonstrate the effects of acceptance cuts,
results are shown first without cuts and then with typical
acceptance cuts imposed to the leptons and the photon.

Figures.~6 and 7 show the $\phi$ and $\cos\theta$ distributions
in the CS and GJ frame without cuts. The curves in Figs.~6
and 7 can be obtained from the results in Fig.~4 using
the coefficients  $\alpha$ and $\beta$ defined in
Eqs.~(\ref{alphadef},\ref{betadef}).
One observes a fairly strong
$\phi$ dependence both in the CS [Fig.~6(a)] and the GJ [Fig.~7(a)]
frame for the three $p_t(\Jpsi)$ bins, which is expected from
$\beta=A_2/4$ and the results in Fig.~4.
The $\phi$ dependence
is larger for high $p_T(\Jpsi)$ bins in the CS frame,
whereas an opposite behavior is found for the GJ frame.
The coefficient $\alpha$ in Eq.~(\ref{alphadef}) is negative for the
three bins in the CS frame and positive in the GJ frame.
The corresponding
$\cos\theta$ distributions are therefore decreasing [increasing] with
$\cos\theta$ in the CS (Fig.~6(b)) [GJ (Fig.~7(b))] frame.
Let us recall, that the angular distributions in Figs.~6 and 7 are
determined by the polarization of the $\Jpsi$ with respect to the $z$-axis
in the CS and GJ frame [see Eq.~(\ref{ghidef})].

Figs.~8 and 9 show the $\phi$ and $\cos\theta$ distributions in the
CS and the GJ frame
for the same bins in $p_T(\Jpsi)$ as in Figs.~6 and 7
but with the cuts
\begin{equation}
p_T(l^-,l^+)>1.8\,\mbox{GeV}\hspace{1cm}
|y(\gamma,l^-,l^+)|<2.5.
\end{equation}
The cuts have a dramatic effect on the shapes of the distribution.
The shapes
of the distributions are now governed by the kinematics of the
surviving events.\footnote{
We thank V.~Barger and R.~J.~N.~Phillips for pointing out this fact.}
The cuts, which are applied in the laboratory
frame, introduce a strong $\phi$  and $\cos\theta$ dependence.
The cuts remove mainly events around $\phi=0,\pi,2\pi$ [Figs.~8,~9~(a)]
and  $\cos\theta>0.5$ for the low $p_T(\Jpsi)$ bin [solid line
in Figs.~8,~9(b)].
The $\phi$ distributions in Figs.~8,~9(a) are  very different from the
``dynamical'' $\phi$ distribution in Figs.~6,~7(a) and there is only very
little sensitivity to the polarization dependence (shown in Figs.~6,~7)
left in Figs.~8 and 9.

We have also analyzed  the effect of the cuts separately by using  the
correct matrix element for $\Jpsi$  production, but with isotropic
decay of the $\Jpsi$, {\it i.e.}, neglecting spin correlations
between $\Jpsi$ production and decay. The angular distributions in this
case are very similar to the ones shown in Figs.~8,~9 for the full matrix
element.

In Figs.~10 and 11 we show ratios of the $\phi$ and $\cos\theta$
distributions in the CS and GJ frame
for the same bins in $p_T(\Jpsi)$ as in Figs.~6--9; the
distribution with full polarization has been divided by the
distribution obtained with isotropic decay of the $\Jpsi$. Cuts
are included in both cases. The large effects from the cuts
are expected to almost cancel in this ratio\footnote{
Analogous ratios have been recently applied in \cite{jim}
to regain the sensitivity to polarization effects in $W$ and $Z$ production
in hadronic collisions.}. In fact, we
nearly recover the $\phi$ and $\cos\theta$ dependence of Figs.~6 and 7
in Figs.~10 and 11.
Especially, the two high $p_T(\Jpsi)$ bins
[dashed and dotted lines in Figs.~10 and 11]
contain most of the polarization dependence seen in Figs.~6 and 7.
Comparing the results in the CS [Figs.~10 and 6] and the
GJ frame [Figs.~11 and 7], the ratios in the  CS frame are more
sensitive to the $\phi$ distribution,
whereas the ratios in the GJ frame are more sensitive to the
$\cos\theta$ distribution, when cuts are applied.

Therefore, to regain sensitivity to the
polarization effects in the presence of large kinematic cuts, we
propose to divide the experimental distributions by the Monte Carlo
distributions obtained using isotropic $\Jpsi$ decay.

\section{SUMMARY}
Hadronic $\Jpsi+\gamma$ production has been evaluated in the
nonrelativistic bound state model.
In leading order this final state can only be produced by gluon fusion.
Analytical formula for
the decay lepton distributions
in terms of four  structure functions are presented.
If no cuts are applied, the angular distribution
in the $\Jpsi$ rest frame is  determined by the polarization of the $\Jpsi$.
We present Monte Carlo studies of the leptonic decay products
of high $p_T$ $\Jpsi$'s produced in association with a photon,
when cuts are imposed on the photon and leptons.
When acceptance cuts are imposed on the leptons, the angular distributions
are dominated by kinematical effects rather than polarization effects.
Polarization effects can be maximized
by minimizing the cuts.
 Alternatively, it
may be possible to retain $\Jpsi$ polarization effects by
``dividing out'' the kinematic effects, {\it i.e.}, if the
histogrammed data are divided by the theoretical result for isotropic
$\Jpsi$  decay.

\bigskip
\noindent{\bf Acknowledgements}
\medskip

We  thank V. Barger, R.J.N. Phillips for insight regarding the effects of the
cuts on the angular distributions.
We thank J. Ohnemus, W.J. Stirling and D. Summers for useful discussions.
The work of CSK is supported in part by the Korean Science and Engineering
Foundation, in part by Non-Direct-Research-Fund, Korea Research Foundation
1993, in part by the Center for Theoretical Physics, Seoul National University,
and in
part by the Basic Science Research Institute Program, Ministry of Education,
1994, Project No. BSRI-94-2425.
The work of EM is supported in part by the U.S. Department of Energy under
contract Nos. DE-AC02-76ER00881 and DE-FG03-91ER40674,  by Texas
National Research Laboratory Grant No.~RGFY93-330, and by the
University of Wisconsin Research Committee with funds granted by the
Wisconsin Alumni Research Foundation.

\newpage
\noindent
\appendix{Structure functions and helicity cross sections}
In this appendix, we will derive the partonic helicity
cross section in Eqs.~(\ref{partondef}-\ref{lastgj})
for the production of polarized $\Jpsi$'s

The  production cross section for
\begin{equation}
g(p_1)+g(p_2)\rightarrow \Jpsi(P)  + \gamma(k)\>,
\end{equation}
is
\begin{equation}
d\hat{\sigma}=\frac{1}{2s} |M|^2 dPS^{(2)}\>,
\hspace{1.5cm}
\mbox{with}
\hspace{1.5cm}
 dPS^{(2)}=\frac{1}{8\pi s}\delta(s+t+u-P^2)\,\, dt du\>.
\end{equation}
The amplitude which describes the coupling of the $\gamma g g$ to the
$\Jpsi$ can be calculated within the bound state formalism
\cite{csm}
to be
\begin{equation}
A^{gg \gamma}= \frac{1}{2} \frac{1}{\sqrt{4\pi M_{\psi}}} R(0)
   \Tr \left[ {\cal O}^{0} (P\slh-M_{\psi}) (-E\slh) \right]\>,
\label{ampli}
\end{equation}
where $R(0)$ denotes the radial wave function of the bound state, which
can be calculated either from potential models or can be
 related to the leptonic
decay rate
\begin{equation}
|R(0)|^{2} = \frac{M_{\psi}^{2}}{4\alpha^{2}(2/3)^{2}} \Gamma_{ee}\>.
\end{equation}
In Eq. (\ref{ampli}) $E$ denotes the $\Jpsi$  polarization.
The amplitude ${\cal O}^{0}$ can be obtained from the amplitude
for the gluon gluon fusion into  a free quark pair at threshold
 and a photon:
\begin{eqnarray}
 {\cal O}^{0} & = & \frac{-1}{4}  \left[
 \frac{{\varepsilon\slh}_{1}(P\slh-2{p\slh}_{1}+m_{\Psi}) {\varepsilon\slh}^{*}
 (-P\slh+2{p\slh}_{2}+m_{\Psi}) {\varepsilon\slh}_{2}}{Pp_{1} \;\; Pp_{2}}
\nonumber \right.\\ & & \;\;\mbox{}
+\frac{{\varepsilon\slh}^{*}(-P\slh+2{p\slh}_{1}+2{p\slh}_{2}+m_{\Psi})
 {\varepsilon\slh}_{1} (-P\slh-2{p\slh}_{2}+m_{\Psi})
 {\varepsilon\slh}_{2}} {(Pp_{1}+Pp_{2}-2p_{1}p_{2}) \;\;Pp_{2}}
\nonumber \\ & & \;\;\left.
 {}+\frac{{\varepsilon\slh}_{1}(P\slh-2{p\slh}_{1}+m_{\Psi})
 {\varepsilon\slh}_{2} (P\slh-2{p\slh}_{1}-2{p\slh}_{2}+m_{\Psi})
 {\varepsilon\slh^{*}}}{Pp_{1} \;\; (Pp_{1}+Pp_{2}-2p_{1}p_{2})}
\right]
  \nonumber \\ & &\;\; {}+[1\leftrightarrow 2]\>.
\end{eqnarray}
Here $p_{1},p_{2},\varepsilon_{1},\varepsilon_{2}$ are the momenta and
polarization vectors of the two gluons and $\varepsilon$ is the polarization
vector of the  photon.
Coupling constants, spin average factors
and color matrices contribute a factor
$ (4\pi\,\alpha_{s}\,e\,2/3)^{2} (1/4)(2/3) $ to production cross section.
After summing over the polarizations
$\varepsilon_{1\,\alpha},\varepsilon_{2\,\beta},\varepsilon_{\mu}$
we define the density matrix elements of the $\Jpsi$ as
\begin{equation}
H^{\sigma\sigma^{\prime}}=
E_{\delta}(\sigma)T^{\delta\delta'}
E_{\delta'}^{\ast}(\sigma^{\prime})\>,
\label{hssdef}
\end{equation}
where the hadronic tensor $T^{\delta\delta'}$ is
\begin{eqnarray}
 T^{\delta\delta'}& = &
\frac{1}{2048 \mpsi^2}\,\,
  \Tr \left[ {\cal
O}^{\alpha\,\mu\,\beta} (P\slh-\mpsi) \gamma^{\delta} \right]
 \Tr \left[ {\cal
O}_{\alpha\;\mu\;\beta} (P\slh-\mpsi) \gamma^{\delta'}
\right]^{*} \label{HJpsi}\>,
\end{eqnarray}
and
  \begin{equation}
    \begin{array}{ll}
E_{\delta}(\pm)&=\frac{1}{\sqrt{2}}\,(0;\pm1,-i,0)\>,\\[3mm]
E_{\delta}(0)  &=(0;0,0,1)\>,
    \end{array}
\hspace{5mm}
\label{polvekdef}
\end{equation}
are the polarization vectors
for the $\Jpsi$ defined with respect to some coordinate axis
in its rest frame (see below).

The angular dependence of the decay leptons from the
$\Jpsi$ can be extracted by introducing the following linear combinations of
the density matrix elements [We follow the technique presented
in Appendix C of \cite{npb}]:
\begin{equation}
    \begin{array}{lllll}
H^{U+L} & = \hspace{3mm}H^{00}+ H^{++} + H^{--}\>, &\hspace{5mm}&
        &
          \\[3mm]
H^{L} & = \hspace{3mm}H^{00}    \>,      & \hspace{5mm}&
        & 
        \\[3mm]
H^{T} & =\hspace{3mm} \frac{1}{2}( H^{+-}+H^{-+})\>,  &\hspace{5mm}&
        &  
        \\[3mm]
H^{I} & = \hspace{3mm}\frac{1}{4}
           \left(H^{+0}+H^{0+}-H^{-0}-H^{0-}\>. \right)&\hspace{5mm}&
      &
    \end{array}
\label{ghidef}
\end{equation}
Collecting all remaining factors, one has
\begin{eqnarray}
\frac{sd\sigma}{dtdu}&=&
\frac{16\pi \alpha\alpha_s^2 \mpsi}{27 s }
|R(0)|^2\,\,
\left\{
(1+\cos^2\theta)\,\, H^{U+L}
+ (1-3\cos^2\theta) \,\,H^L \right. \nonumber\\
&&\left.  + 2\sqrt{2} \sin 2\theta \cos\phi\,\, H^I
+2 \sin^{2}\theta\cos 2\phi \,\,H^T \right\}\delta(s+t+u-P^2)\>.
\label{anga}
\end{eqnarray}
In Eq. (\ref{anga})
$\theta$ and $\phi$ denotes the lepton angles in the lepton pair
rest frame.
What remains is the calculation of the $H^{\alpha}$ in the
CS and GJ frame.

For this purpose we
choose a specific
representation for $E_{\delta}(\sigma)E_{\delta'}^{\ast}(\sigma')$
in Eq. (\ref{hssdef})
and express them in terms of
of $g_{\delta\delta'},
P_{\delta}P_{\delta'}, p_{i\,\delta}p_{j\,\delta'}\,\,\,\,(i,j=1,2)$.
Let us define the following covariant projections:
\begin{equation}
\tilde{T}^{\beta}\equiv {\cal{P}}_{\delta\delta'}^{\beta}\,
H^{\delta\delta'},\hspace{3mm}
(\beta \in
\{U+L,L_1,L_2,L_{12}\})
\label{tildet}
\end{equation}
where
\begin{equation}
    \begin{array}{ll}
{\cal{P}}_{\delta\delta'}^{U+L}&=\,-\hat{g}_{\delta\delta'}\>,
        \\[3mm]
{\cal{P}}_{\delta\delta'}^{L_{1}}&=\,\frac{1}{\hat{E}_1^2}\,\,
\hat{p}_{1\delta}\hat{p}_{1\delta'}\>,\\[3mm]
{\cal{P}}_{\delta\delta'}^{L_{2}}&=\,\frac{1}{\hat{E}_2^2}\,\,
\hat{p}_{2\delta}\hat{p}_{2\delta'}\>,\\[3mm]
{\cal{P}}_{\delta\delta'}^{L_{12}}&=\,\frac{1}{\hat{E}_1\hat{E}_2}\,\,
(\hat{p}_{1\delta}\hat{p}_{2\delta'} +
\hat{p}_{1\delta'}\hat{p}_{2\delta} )\>.\\[3mm]
        \end{array}
\label{projektorpc}
  \end{equation}
We have introduced the hatted tensors
\begin{equation}
\hat{g}_{\delta\delta'}= {g}_{\delta\delta'} -
\frac{P_{\delta}P_{\delta'}}{P^2}\>,
\hspace{1cm}
\hat{p}_{i,\delta}=p_{i\,\delta}-\frac{p_iP}{P^2}p_{i\,\delta}\>,
\hspace{1cm}
\hat{E}_1=\frac{P^2-t}{2\sqrt{P^2}}
\hspace{1cm} \hat{E}_2=\frac{P^2-u}{2\sqrt{P^2}}\>.
\end{equation}
The helicity structure functions $H^{\alpha}$ in Eq.~(\ref{anga})
for a given $\Jpsi$ rest frame
are linear combinations of the covariant projections
in Eq.~(\ref{tildet}).

For the CS frame one has \cite{npb}:
\begin{equation}
\left(
\begin{array}{c}
H^{U+L}\\[3mm]
H^L\\[3mm]
H^T\\[3mm]
H^I\\[3mm]
\end{array}
\right)_{CS}
=
\left(
    \begin{array}{cccc}
1   &  0  & 0 &  0 \\[2mm]
0   &  \frac{1}{4\,\cos^{2}\gamma_{CS}}
         &  \frac{1}{4\,\cos^{2}\gamma_{CS}}
         &  \frac{-1}{4\,\cos^{2}\gamma_{CS}}
          \\[3mm]
\frac{1}{2}
         &   -\frac{(1+\cos^{2}\gamma_{CS})}
                 {8\,\sin^{2}\gamma_{CS}\cos^{2}\gamma_{CS}}
         &   -\frac{(1+\cos^{2}\gamma_{CS})}
                 {8\,\sin^{2}\gamma_{CS}\cos^{2}\gamma_{CS}}
        &   \frac{(1-3\cos^{2}\gamma_{CS})}
            {8\,\sin^{2}\gamma_{CS}\cos^{2}\gamma_{CS}}
                        \\[3mm]
0        & \frac{1}{4\sqrt{2}\,\sin\gamma_{CS}\cos\gamma_{CS}}
        & \frac{-1}{4\sqrt{2}\,\sin\gamma_{CS}\cos\gamma_{CS}}
        & 0   \\
    \end{array}
\right)
\
\left(
\begin{array}{c}
H^{U+L}\\[3mm]
H^{L_1}\\[3mm]
H^{L_2}\\[3mm]
H^{L_{12}}\\[3mm]
\end{array}
\right)
\label{csmatrix}
\end{equation}
where
\begin{equation}
\cos\gamma_{CS}=
\sqrt{\frac{\mpsi^2 s}{(t-\mpsi^2)(u-\mpsi^2)}}\>,
\hspace{2cm}
\sin\gamma_{CS}=-\sqrt{1-\cos^2\gamma_{CS}}\>.
\end{equation}
The explicit results for $H^{\alpha}_{CS}$ are given in
Eqs.~(\ref{firstcs}-\ref{lastcs}).

The results for the  GJ frame
can be obtained from:
\begin{equation}
\left(
\begin{array}{c}
H^{U+L}\\[3mm]
H^L\\[3mm]
H^T\\[3mm]
H^I\\[3mm]
\end{array}
\right)_{GJ}
=
\left(
    \begin{array}{cccc}
1   &  0  & 0 &  0 \\[2mm]
0   &  1  & 0 &  0 \\[3mm]
\frac{1}{2}
         &   -\frac{(1+\cos^{2}\gamma_{GJ})}
                 {2\sin^2\gamma_{GJ}}
         &   -\frac{1}{\sin^2\gamma_{GJ}}
        &   \frac{\cos\gamma_{GJ}}
            {\sin^2\gamma_{GJ}}
                        \\[3mm]
0        & \frac{-\cos\gamma_{GJ}}{\sqrt{2}\,\sin\gamma_{GJ}}
        & 0
        &  \frac{1}{2\sqrt{2}\,\sin\gamma_{GJ}}   \\
    \end{array}
\right)
\
\left(
\begin{array}{c}
H^{U+L}\\[3mm]
H^{L_1}\\[3mm]
H^{L_2}\\[3mm]
H^{L_{12}}\\[3mm]
\end{array}
\right)
\label{GJmatrix}
\end{equation}
and
\begin{equation}
\cos\gamma_{GJ}=
 {1-\frac{2\mpsi^2 s}{(t-\mpsi^2)(u-\mpsi^2)}}\>,
\hspace{2cm}
\sin\gamma_{GJ}=-\sqrt{1-\cos^2\gamma_{GJ}}\>.
\end{equation}
The explicit results for $H^{\alpha}_{GJ}$ are given in
Eqs.~(\ref{firstgj}-\ref{lastgj}).

\appendix{LORENTZ TRANSFORMATIONS TO THE CS AND GJ FRAME}
For completeness, we list here the transformation matrices to the
CS and the GJ frame.
In the laboratory frame, the $z$-direction is defined by the proton
momentum and the $x$-direction is defined by the transverse momentum
of the $\Jpsi$. The Lorentz transformation matrix for the  boost
from the laboratory frame to the CS  and GJ frame is given by
[$P$ denotes the $\Jpsi$ momentum]
\begin{equation}
\left(
\begin{array}{c}
E\\[3mm]
p_x\\[3mm]
p_y\\[3mm]
p_z\\[3mm]
\end{array}
\right)_{CS}
= \
\left(
\begin{array}{cccc}
\frac{P_0}{\sqrt{P^2}}\hspace{3mm} & -\frac{p_T^{}}{\sqrt{P^2}}   &
\hspace{3mm}0\hspace{3mm}       & -\frac{P_3}{\sqrt{P^2}}        \\[3mm]
-\frac{p_T^{}\, P_0}{\sqrt{P^2}\,X_T} & \frac{X_T}{\sqrt{P^2}}    &
0                      & \frac{p_T^{}\,P_3}{\sqrt{P^2}\,X_T}\\[3mm]
0                      & 0                                     &
1                      & 0                               \\[3mm]
-\frac{P_3}{X_T}       & 0                                     &
0                      & \frac{P_0}{X_T}\\[3mm]
\end{array}
\right)
\
\left(
\begin{array}{c}
E\\[3mm]
p_x\\[3mm]
p_y\\[3mm]
p_z\\[3mm]
\end{array}
\right)_{lab}
\label{EQ:BOOST}
\end{equation}
\begin{equation}
\left(
\begin{array}{c}
E\\[3mm]
p_x\\[3mm]
p_y\\[3mm]
p_z\\[3mm]
\end{array}
\right)_{GJ}
= \
\left(
\begin{array}{cccc}
\frac{P_0}{\sqrt{P^2}}\hspace{3mm} & -\frac{p_T^{}}{\sqrt{P^2}}   &
\hspace{3mm}0\hspace{3mm}       & -\frac{P_3}{\sqrt{P^2}}        \\[3mm]
-\frac{p_T^{}\, (P_0-P_3)}{X_T^2}&  1    &
0                      & -\frac{p_T^{}\,(P_0-P_3)}{X_T^2}\\[3mm]
0                      & 0                                     &
1                      & 0                               \\[3mm]
-\frac{p_T^2P_0+P^2P_3}{\sqrt{P^2}X_T^2}       & \frac{p_T}{\sqrt{P^2}}
                                     &
0                      & \frac{P^2 P_0+p_T^2 P_3}{\sqrt{P^2}{X_T^2}}\\[3mm]
\end{array}
\right)
\left(
\begin{array}{c}
E\\[3mm]
p_x\\[3mm]
p_y\\[3mm]
p_z\\[3mm]
\end{array}
\right)_{lab}
\label{EQ:GJBOOST}
\end{equation}
where
\begin{equation}
X_T=\sqrt{P^2+p_T^2} \>.
\end{equation}
\newpage
\noindent
{\bf Figure captions}
\begin{itemize}
\item[{\bf Fig. 1}]
The transverse momentum distribution of the $\Jpsi$ [or $\gamma$]
in $\Jpsi+\gamma$ production at the Tevatron collider center of mass
energy [$\sqrt{S}=1.8$ TeV]
with subsequent $\Jpsi\rightarrow 
\mu^-\mu^+$ decay.
The following cuts have been applied: $|y(\gamma,l^-,l^+)|<2.5$ and
$p_T(l^-,l^+)>1.8$ GeV.
The renormalization and factorization scales  $\mu^2$ are set
equal to
$(p_{_T}^2(\gamma)+\mpsi^2)$ [solid],
$1/4\,(p_{_T}^2(\gamma)+\mpsi^2)$ [dashed] and
$4\,(p_{_T}^2(\gamma)+\mpsi^2)$ [dotted].
\item[{\bf Fig. 2}]
The normalized rapidity distribution of the $\gamma$
in $\Jpsi+\gamma$ production
at the Tevatron collider center of mass
energy [$\sqrt{S}=1.8$ TeV]
with subsequent $\Jpsi\rightarrow \mu^-\mu^+$ decay.
The following cuts have been applied:
$p_T(\gamma,l^-,l^+)>3$ GeV.
\item[{\bf Fig. 3}]
The transverse momentum distribution of the leptons  from
the decay of the $\Jpsi$
in $\Jpsi+\gamma$ production  at the
Tevatron collider center of mass
energy [$\sqrt{S}=1.8$ TeV].
The following cuts have been applied:
$|y(\gamma,l^-,l^+)|<2.5, \,\,p_T(l^-,l^+)>1.8$ GeV
and $p_T(\gamma)> 2$ GeV.
Results are shown for the positive (negative)
charged lepton [dashed lines]
as well as the distribution for the lepton
with bigger (subscript $b$) and smaller
(subscript $s$) $p_{_T}$ without identifying the charge.
\item[{\bf Fig. 4}]
Angular coefficients $A_0, A_1$ and $A_2$ for $\Jpsi+\gamma$
production and decay in the
CS frame (a) and GJ frame (b)
as a function of the $\Jpsi$ transverse momentum
at $\sqrt{S}=1.8$ TeV.
No cuts have been applied.
\item[{\bf Fig. 5}]
Same as Fig.~4 for the $y(\Jpsi)$ distribution.
\item[{\bf Fig. 6}]
a) Normalized $\phi$ and b) normalized $\cos\theta$ distributions
of the  leptons from $\Jpsi$ decay in the CS frame.
Results are shown for three bins in $p_T^{}(\Jpsi)$:\\
2~GeV $<\, p_T^{}(\Jpsi)\, < 5$~GeV [solid],\\
5~GeV $<\, p_T^{}(\Jpsi)\, < 8$~GeV [dashed],\\
8~GeV $<\, p_T^{}(\Jpsi)$   [dotted].\\
No cuts  have been applied.
\item[{\bf Fig. 7}]
Same as Fig.~6 in the GJ frame.
\item[{\bf Fig. 8}]
Same as Fig.~6 but with  the cuts
$
|y(\gamma,l^-,l^+)|<2.5,\>
p_T(l^-,l^+)>1.8\,\mbox{GeV}.
$
\item[{\bf Fig. 9}]
Same as Fig.~7 but with  the cuts
$
|y(\gamma,l^-,l^+)|<2.5,\>
p_T(l^-,l^+)>1.8\,\mbox{GeV}
$
\item[{\bf Fig. 10}]
Ratios of distributions obtained with full polarization effects to
those obtained with isotropic decay of the $\Jpsi$.
Parts a) and b)
are the ratios for the $\phi$ and $\cos\theta$ distributions in the
CS frame, respectively. The cuts
$
|y(\gamma,l^-,l^+)|<2.5,\>
p_T(l^-,l^+)>1.8\,\mbox{GeV}
$
are included.
\item[{\bf Fig. 11}]
Same as Fig.~10 for the GJ frame.
\end{itemize}

\def\npb#1#2#3{{\it Nucl. Phys. }{\bf B #1} (#2) #3}
\def\plb#1#2#3{{\it Phys. Lett. }{\bf B #1} (#2) #3}
\def\prd#1#2#3{{\it Phys. Rev. }{\bf D #1} (#2) #3}
\def\prl#1#2#3{{\it Phys. Rev. Lett. }{\bf #1} (#2) #3}
\def\prc#1#2#3{{\it Phys. Reports }{\bf C #1} (#2) #3}
\def\pr#1#2#3{{\it Phys. Reports }{\bf #1} (#2) #3}
\def\zpc#1#2#3{{\it Z. Phys. }{\bf C #1} (#2) #3}
\def\ptp#1#2#3{{\it Prog.~Theor.~Phys.~}{\bf #1} (#2) #3}
\def\nca#1#2#3{{\it Nouvo~Cim.~}{\bf A #1} (#2) #3}
\newpage
\sloppy
\raggedright

\end{document}